\documentclass[acmsmall]{acmart}
\AtBeginDocument{%
  }

\setcopyright{acmlicensed}
\acmDOI{XXXXXXX.XXXXXXX}
\acmISBN{978-1-4503-XXXX-X/2018/06}

\usepackage{subfigure} 
\usepackage{tcolorbox}
\usepackage{todonotes}

\begin{document}

\title{Semantic Intelligence: A Bio-Inspired Cognitive Framework for Embodied Agents}

\author{Wenbing Tang}
\email{wenbing.tang@ntu.edu.sg}
\affiliation{%
  \institution{Nanyang Technological University}
  \country{Singapore}
}

\author{Meilin Zhu}
% \authornotemark[2]
\email{zhuml@ios.ac.cn}
\affiliation{%
  \institution{University of Chinese Academy of Sciences}
  \country{China}
}

\author{Fenghua Wu}
\email{fenghua.wu@ntu.edu.sg}
\affiliation{%
  \institution{Nanyang Technological University}
  \country{Singapore}
}

\author{Yang Liu}
\authornotemark[1]
\email{fenghua.wu@ntu.edu.sg}
\affiliation{%
  \institution{Nanyang Technological University}
  \country{Singapore}
}

% \affiliation{%

% \affiliation{%

% \affiliation{%

% \affiliation{%

% \affiliation{%

% \affiliation{%

% \affiliation{%

% %%
% %% By default, the full list of authors will be used in the page
% %% headers. Often, this list is too long, and will overlap
% %% other information printed in the page headers. This command allows
% %% the author to define a more concise list
% %% of authors' names for this purpose.

\begin{abstract}
Recent advancements in Large Language Models (LLMs) have greatly enhanced natural language understanding and content generation. However, these models primarily operate in disembodied digital environments and lack interaction with the physical world. To address this limitation, Embodied Artificial Intelligence (EAI) has emerged, focusing on agents that can perceive and interact with their surroundings. Despite progress, current embodied agents face challenges in unstructured real-world environments due to insufficient semantic intelligence, which is critical for understanding and reasoning about complex tasks. This paper introduces the Semantic Intelligence-Driven Embodied (SIDE) agent framework, which integrates a hierarchical semantic cognition architecture with a semantic-driven decision-making process. This enables agents to reason about and interact with the physical world in a contextually adaptive manner. The framework is inspired by biological cognitive mechanisms and utilizes bio-inspired principles to design a semantic cognitive architecture that mimics how humans and animals integrate and process sensory information. We present this framework as a step toward developing more intelligent and versatile embodied agents.
\end{abstract}





\maketitle

\section{Introduction}
In recent years, Large Language Models (LLMs) have revolutionized artificial intelligence (AI) by demonstrating unprecedented capabilities in natural language understanding, multimodal reasoning, and content generation. 
As technology has advanced, LLMs have further evolved into LLM-based agents capable of performing complex reasoning tasks, planning multi-step action sequences, and interacting with various APIs and tools through the seamless integration of language understanding with actionable capabilities~\cite{liu2025advances,yehudai2025survey}. Despite these impressive advancements, both LLMs and LLM-based agents fundamentally operate within disembodied digital environments~\cite{liu2025embodied,wu2024avatar,sonlu2025effects}. They interact with structured data, software systems, or virtual simulations, but lack direct perception of, and action within, the physical world.

To bridge the gap between digital reasoning and physical reality, the field of Embodied Artificial Intelligence (EAI) has emerged as a critical research frontier. EAI fundamentally shifts the focus from purely computational intelligence to creating agents, such as robots, that possess a physical body~\cite{liu2025embodied,xing2025towards}. These agents are able to directly perceive, interact with, and learn from their surrounding environment through sensorimotor loops. This paradigm moves beyond disembodied algorithms, which process abstract data, to systems that engage in physical manipulation~\cite{li2024manipllm}, navigation~\cite{zheng2024towards}, and dynamic interaction within complex, unstructured settings. The core tenet of EAI is that intelligence is deeply intertwined with physical experience.
As a result, EAI is widely regarded as a crucial pathway toward achieving Artificial General Intelligence (AGI), as it challenges intelligence to confront the complexities and uncertainties inherent in the real world. In recent years, LLMs have emerged as powerful tools for building embodied agents~\cite{li2024embodied,yang2025embodiedbench}.

\begin{figure}[htbp]
    \centering
    \subfigure[An autonomous driving scenario.]{
        \includegraphics[width=0.55\textwidth]{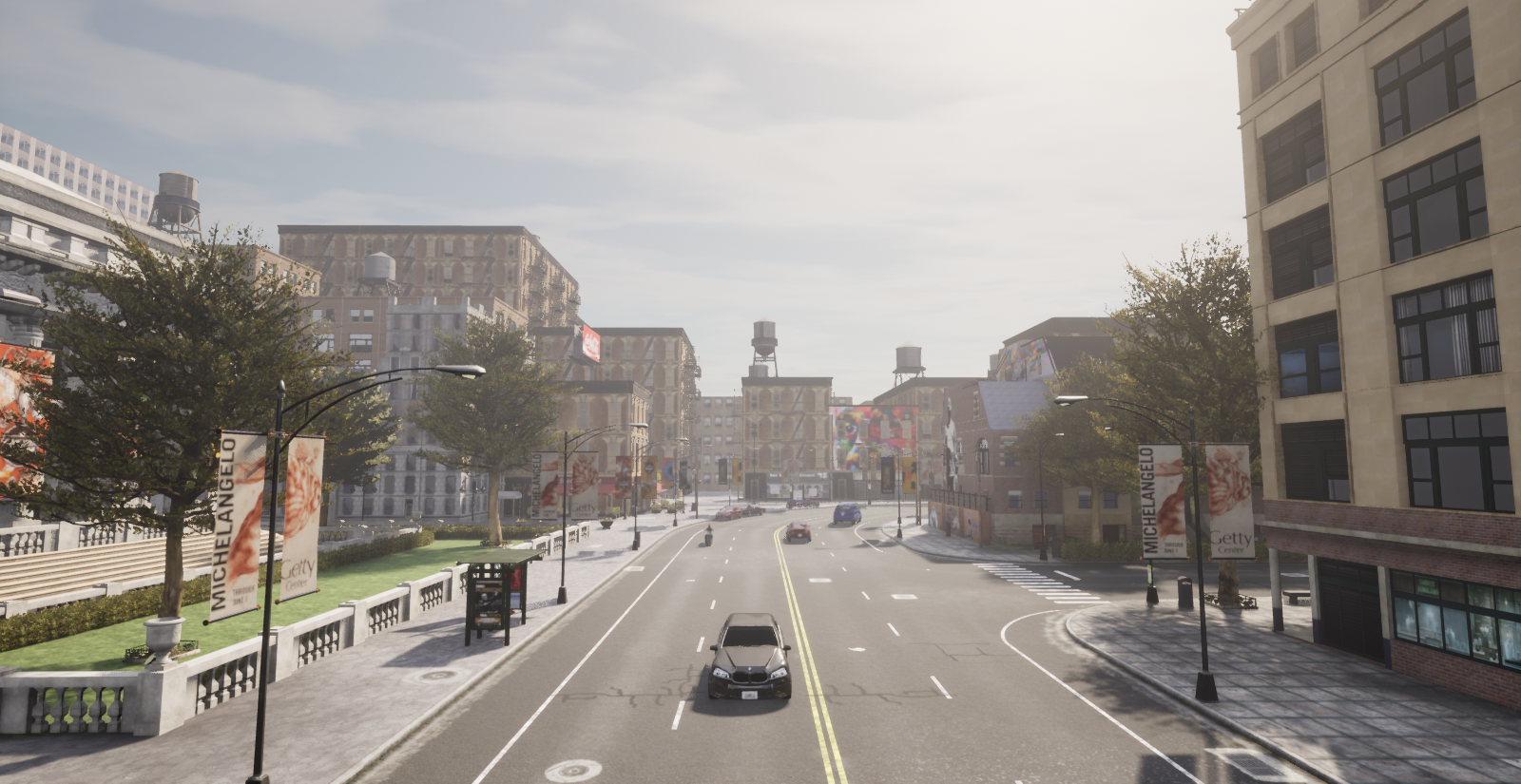}
    }
    \hfill
    \subfigure[A scenario for embodied task planning.]{
        \includegraphics[width=0.38\textwidth]{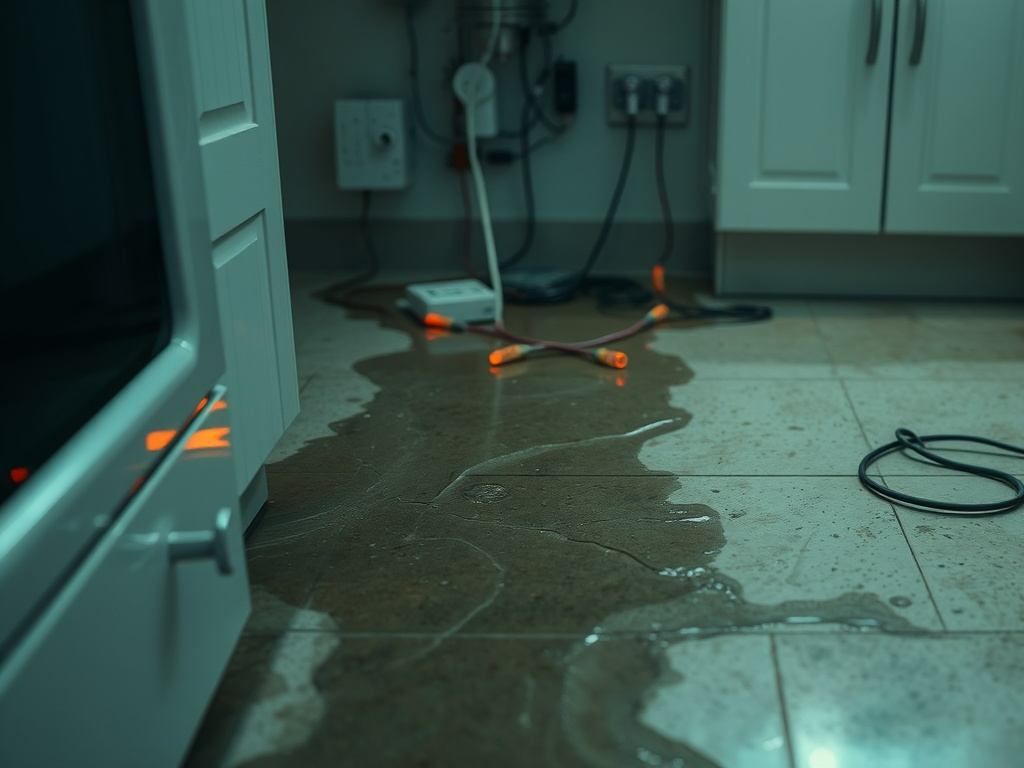}
    }
    \caption{Two scenarios demonstrating the necessity of semantic understanding in embodied agents.}
    \label{motivating-examples}
\end{figure}

% Despite rapid progress in both LLMs and robotic systems, several critical limitations persist in the development of semantically intelligent embodied agents. 
Current approaches to embodied AI demonstrate promising capabilities in controlled settings but exhibit significant performance degradation when deployed in open-world environments~\cite{liu2024aligning}. 
For instance, as shown in Fig.~\ref{motivating-examples}(a), an autonomous driving agent should focus on semantically relevant entities
such as pedestrians, vehicles, and traffic lights, rather than 
irrelevant visual details like building shapes or colors.
% For instance, as shown in Fig.~\ref{motivating-examples}(a), a household agent that successfully identifies and retrieves specific objects (with instruction \textit{"bring me the black mug from the kitchen counter"}) in a laboratory environment with consistent lighting, predefined object positions, and limited distractors often fails when the same task is performed in a real home setting where the cup may be partially occluded, oriented differently, or subject to varying lighting conditions.
Another example is illustrated in Fig.~\ref{motivating-examples}(b), where the instruction is: \textit{“Fix the water leakage on the kitchen floor.”}
In this scenario, water is leaking onto the floor and is about to reach nearby electrical wires. While the robot attempts to fix the leak directly, a human would recognize the potential danger of electrical conduction and would first cut off the power supply before addressing the water issue.
Importantly, these failures are not merely the result of perceptual noise or insufficient training data. They reveal a more fundamental limitation: \textbf{\textit{current embodied agents lack genuine semantic intelligence, namely the capacity to extract, represent, reason, and apply about semantics.}}
This ``semantic gap'' hinders agents from performing tasks that require common sense, flexible adaptation, or nuanced interaction. An agent might recognize an object as a ``mug'' but fail to understand its function (holding liquid), its typical location (kitchen counter), or its state (empty vs. full, upright vs. sideways) without explicit programming or extensive, task-specific training. 

To truly empower embodied agents to operate intelligently and autonomously alongside humans, we must equip them with the capacity to build and utilize rich semantic knowledge of the physical world. 
However, building semantically intelligent embodied agents requires addressing several key challenges:
(1) \textit{Semantic perception} in the physical world remains difficult due to partial observability, sensor noise, dynamic object appearances, and ambiguous object categories~\cite{zhu2020dark,henson2012semantic}. For example, determining whether an object is a “mug” or a “cup” may require understanding subtle visual cues, contextual usage, or even user preferences, which go beyond traditional object detection.
(2) \textit{Semantic extraction and representation} is non-trivial, as it involves converting raw sensory data into high-level, actionable representations~\cite{TraftonHHTKS13}. For instance, when an agent is asked to “\textit{bring me the cup on the table,}” it must not only detect the cup but also understand its position relative to other objects, the table, and the environment, which can be challenging in cluttered or dynamic scenarios.
(3) \textit{Integrating semantic knowledge with low-level control} is another significant challenge. While robots can perform physical tasks like moving or grasping, they often fail to apply semantic reasoning to low-level actions~\cite{li2024embodied}. For example, when an agent is tasked with picking up a fragile object like a glass, it needs to adjust its grip strength and movement based on its understanding of fragility~\cite{li2024multi}, which requires a combination of semantic understanding and precise motor control.
These challenges are crucial obstacles for realizing truly intelligent, adaptable embodied agents that can effectively interact with and understand the physical world.

In this paper, we propose a novel framework for empowering embodied agents with semantic intelligence by integrating a hierarchical semantic cognitive architecture into their decision-making loop. First, inspired by human cognitive mechanisms studied in cognitive science, we emphasize the importance of bridging embodied agents with cognitive theories and propose the concept of \textit{bio-inspired semantic intelligence}. Second, we design \textit{a computable semantic cognitive architecture} that enables agents to perceive, interpret, and reason about the physical world semantically. 
This architecture decomposes semantic cognition into multiple layers, enabling the agent to hierarchically extract high-level cognitive understanding from low-level perception signals.
Third, we introduce \textit{a semantic-driven decision process} that explicitly embeds semantic knowledge into the agent’s decision-making pipeline. 
This process maps task specifications to executable actions while maintaining semantic grounding throughout, ensuring that decisions and behaviors are guided by an understanding of the deeper meaning embedded in physical contexts. 
Finally, we unify the semantic cognitive architecture with the semantic-driven decision process, resulting in a comprehensive framework: \textit{the Semantic Intelligence-Driven Embodied (SIDE) agent framework}.

The primary contribution of this work lies in the proposal of a novel SIDE agent framework, which integrates advanced theories from cognitive science, AI, and robotics. This framework provides both a conceptual foundation and a practical blueprint for developing embodied agents with semantic understanding, offering a tangible solution to the challenge of achieving semantically intelligent agents. Specifically, it underscores the crucial role of simulating human cognitive processes as a means to enable semantic intelligence. We argue that this architecture facilitates more flexible planning, robust execution, and interpretable behavior, distinguishing it from current end-to-end or purely reactive systems.
This paper systematically discusses the key challenges in the development of semantically intelligent agents, proposes a feasible solution through the SIDE framework, and outlines potential applications and future research directions.

% , as illustrated in Figure [Reference to User's Figure - User to Add]. Our approach posits that robust intelligence emerges from a structured process that transforms raw sensory input into meaningful semantic primitives, reasons over these primitives to infer higher-level conceptual, spatial, and temporal knowledge, and leverages this understanding within a goal-directed decision-making process modulated by memory and metacognitive oversight. The architecture explicitly models the aggregation of semantic information, from basic perceptual features (e.g., shape, location, duration) through intermediate reasoning (e.g., arrangement, sequence, function inference) to integrated cognitive understanding.

% In the pursuit of realizing artificial general intelligence (AGI), the importance of
% embodied artificial intelligence (EAI) becomes increasingly apparent.
% Despite advances in EAI, agent reasoning systems still
% struggle to capture the fundamental semantic information that
% humans naturally use to understand and interact with their environment.

% Semantic Cognition: Semantic Memory and Semantic Control

% ACT-R

% Semantic Memory: Factual knowledge about the world, including concepts, words, and their
% relationships [182]. Examples include recalling the meaning of vocabulary terms or knowing the
% capital city of a country.

% Neural Memory Networks
% MEMORY NETWORKS

% Connect agents with cognitive science

\section{Bio-inspired Semantic Intelligence}

The development of semantically intelligent embodied agents has been hindered by fundamental limitations in traditional AI architectures, such as pattern recognition and statistical learning.
Despite advances in visual perception, language processing, and robotics, embodied agents continue to struggle with semantic  understanding in real-world environments~\cite{raychaudhuri2025semantic}.
These limitations underscore the need for a paradigm shift in the conceptualization and implementation of semantic intelligence.

Building semantically intelligent embodied agents requires a theoretical foundation that addresses how semantic knowledge is acquired, represented, and applied in complex physical environments. 
Cognitive science, encompassing disciplines such as psychology, neuroscience, linguistics, and anthropology, offers critical insights into the mechanisms underlying human perception, reasoning, memory, and learning~\cite{friedenberg2021cognitive,cross2021mind}, which are fundamental components of semantic understanding. 
Researchers in cognitive science have systematically investigated a wide range of cognitive functions and mechanisms in both humans and animals~\cite{aggelopoulos2015perceptual,barrett2008evolved,langley2013central,talmi2013enhanced,zhu2024capturing}, encompassing perception, attention, memory, learning, and emotion.
For example, Treisman's Feature Integration Theory (FIT)~\cite{aggelopoulos2015perceptual}, which explains how humans integrate disparate low-level visual features (color, shape, orientation) into coherent object representations through attentional mechanisms.
By implementing computational architectures inspired by FIT's two-stage processing model, we can develop agents with more human-like capabilities for object perception and scene understanding.

Drawing upon these cognitive insights, we propose the concept of \textit{bio-inspired semantic intelligence} as a foundational approach for developing advanced embodied agents:
\begin{center}
\begin{tcolorbox}[colback=blue!5!white,colframe=blue!55!black,width=0.98\textwidth,title={Definition of Bio-inspired Semantic Intelligence}]
{
Bio-inspired Semantic Intelligence refers to the capacity of an agent to achieve semantic understanding by drawing on biological cognitive mechanisms for perceiving, representing, and reasoning about the physical world, ultimately exhibiting human-like contextual awareness, adaptability, and goal-directed behavior.
}
\end{tcolorbox}
\end{center}

By adopting a bio-inspired methodology, we can progress toward developing embodied agents that not only execute tasks but also understand the semantic structure of their operational context. This approach promises to yield systems that are more flexible, interpretable, and capable of collaborating seamlessly with humans, ultimately bridging the gap between artificial computation and true real-world comprehension.

% From a cognitive science perspective, semantic intelligence is deeply tied to how agents integrate sensory data with prior knowledge, contextual understanding, and the ability to reason about cause-and-effect relationships. The human brain, for example, continually integrates incoming sensory information with pre-existing mental representations, drawing on experience, memory, and understanding of the world to form coherent, actionable semantic representations. This process is dynamic and context-dependent, much like the challenges embodied agents face in understanding the complexities of the real world.

% Given this, we propose that bio-inspired semantic intelligence is a promising approach to equipping embodied agents with a level of understanding and reasoning that mimics human cognitive abilities. By integrating cognitive science theories into the design of embodied agents, we aim to create systems that not only perceive the world but understand it in a way that is meaningful and contextually appropriate.

% Thus, we define bio-inspired semantic intelligence as the ability of an embodied agent to perceive, reason, and act based on semantic knowledge derived from a synthesis of sensory input and contextual understanding, inspired by the cognitive processes of biological systems. This approach integrates the perceptual, reasoning, and decision-making capabilities of the agent into a coherent, semantically aware system, drawing inspiration from the cognitive mechanisms that underlie human intelligence.

% \section{Semantic Intelligence-Driven Agent Framework}

\section{Semantic-Intelligence-Driven Embodied (SIDE) Agent Framework}

\begin{figure*}[h]
\includegraphics[width=\textwidth]{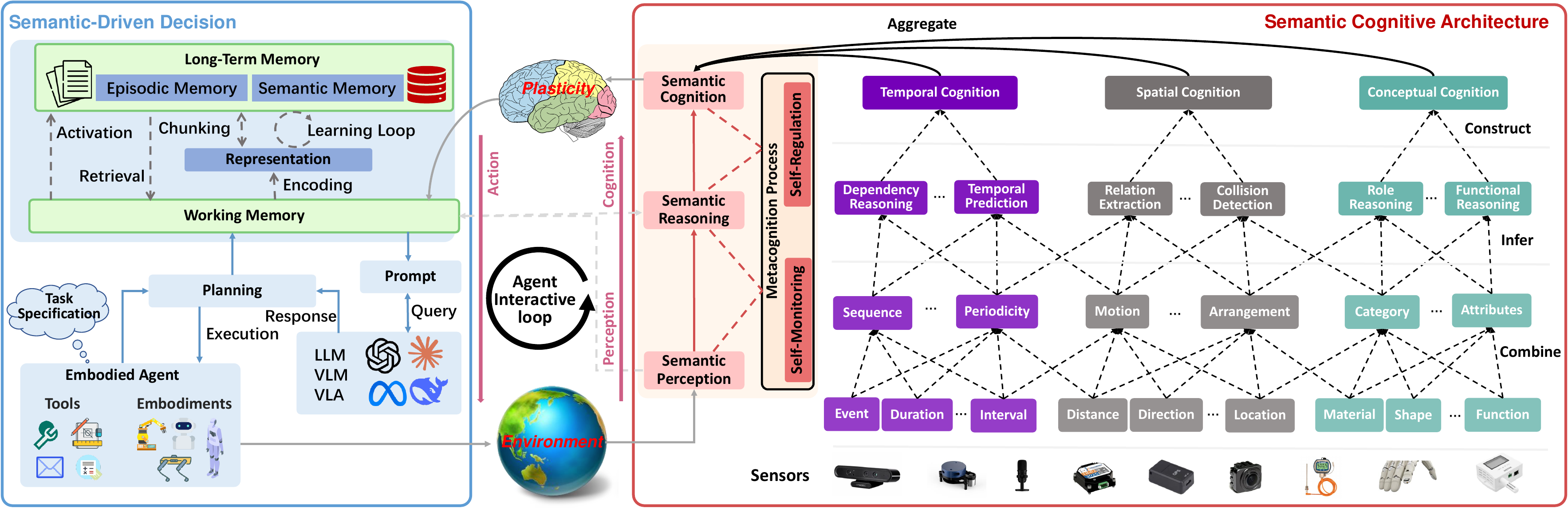}
  \caption{The framework of semantic intelligence-driven embodied agents.}
  % \vspace{-15pt}
  \label{fig-framework}
\end{figure*}

Grounded in the principles of Bio-inspired Semantic Intelligence, this section presents the design of the Semantic Intelligence-Driven Embodied (SIDE) agent framework.
Fig.~\ref{fig-framework} illustrates our proposed solution for equipping embodied agents with semantic intelligence.
At a high level, the SIDE agent framework operates through a closed \textit{perception–cognition-action loop}: as the agent interacts with the environment, perceptual inputs continuously update the semantic cognition module, which in turn reshapes the agent’s planning and behavior.
At its core, the framework consists of two components: the \textit{Semantic Cognitive Architecture} (right side), responsible for building semantic knowledge, and the \textit{Semantic-Driven Decision Loop} (left side), which uses this knowledge to guide planning and action execution.

The semantic cognitive architecture models the agent’s semantic cognition as a hierarchical process. 
It begins with semantic perception, where the agent gathers multimodal information through its sensors. This perceptual data is then processed through a structured reasoning mechanism to infer more complex relationships and concepts.
The highest layer aggregates insights from the reasoning layer into a cohesive, high-level semantic cognition.
Specifically, the entire cognitive architecture is orchestrated by a metacognitive module that monitors and regulates cognitive processes.
The semantic-driven decision loop begins with the interpretation of a task specification, which triggers the planning process. During planning, the agent relies on working memory, which dynamically interacts with long-term memory to retrieve relevant semantic knowledge stored in semantic memory. Working memory provides the immediate context, including the current task, recent perceptions, and relevant semantic knowledge, which is then used to formulate a specific query for the LLM (or VLM, VLA) during planning. The LLM’s response, based on the query, is processed by the planning module and translated into actionable instructions for the agent’s execution.
Together, these components enable the agent to not only perceive and act but also to do so with a continuously updated, contextually grounded understanding of semantic meaning. 
The detailed descriptions of each module are provided in the following sections.

% In this section, we describe the design of the SIDE agent framework.
% The proposed framework, as shown in Fig.~\ref{fig-framework}, presents an integrated architecture designed to empower embodied agents with semantic intelligence.

\subsection{Semantic Cognitive Architecture}

The semantic cognitive architecture constitutes the core semantic knowledge acquisition and representation system within our SIDE agent framework, as illustrated in the right panel of Fig.~\ref{fig-framework}.
Drawing inspiration from hierarchical models of human semantic cognition~\cite{ralph2017neural,pulvermuller2016brain}, the architecture implements a multi-layered processing hierarchy that transforms sensor data into semantic cognition. 

\begin{figure}[h]
\includegraphics[width=0.5\textwidth]{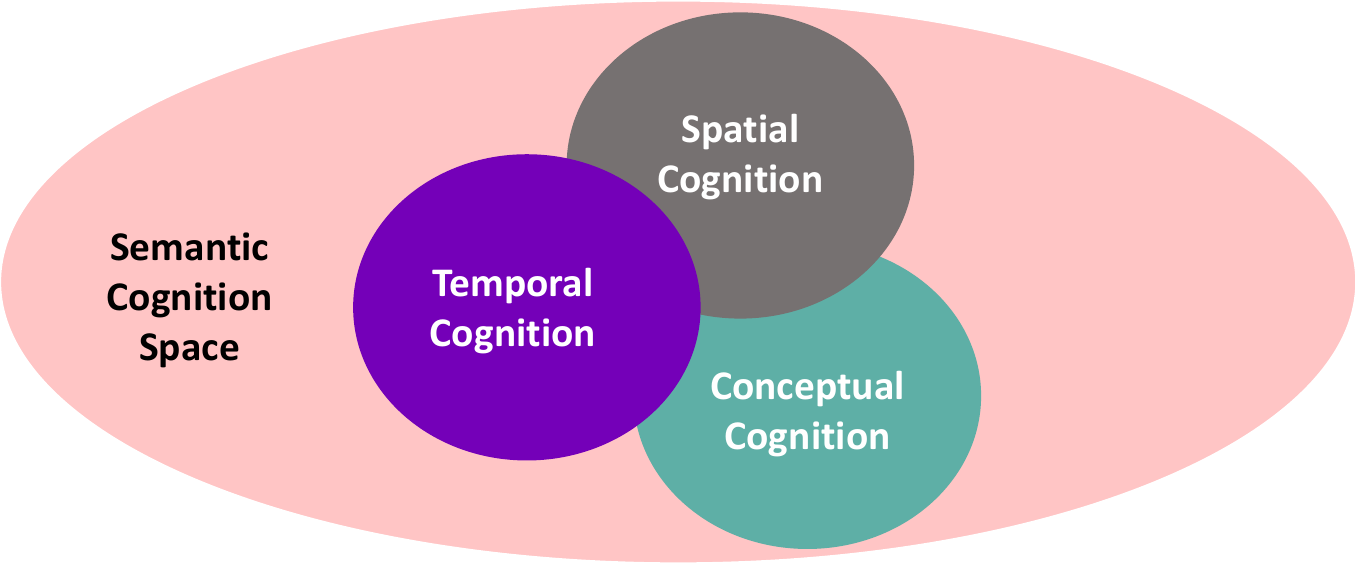}
  \caption{Three basic types of cognition make up semantic cognition.}
  \label{three-basic-cognition}
  % \vspace{-10pt}
\end{figure}

As shown in Fig.~\ref{fig-framework}, in our framework design, we believe that semantic cognition includes three key types of cognition: temporal cognition  (shown in purple), spatial cognition (shown in gray), and conceptual cognition (shown in teal).
This design choice reflects neurobiological evidence suggesting that human semantic processing arises from the integration of different fundamental cognitive components\cite{ralph2017neural}.
Specifically, temporal cognition allows the agent to perceive and reason about time-related relationships, such as event sequences, causal dependencies, and temporal patterns. This capability supports the understanding of narrative structures, anticipation of future states, and interpretation of the sequential nature of actions and events.
Spatial cognition supports the agent’s understanding of spatial layouts, geometric relations, and orientation, which are essential for navigation, object localization, and physical interaction with the environment. 
Conceptual cognition refers to the agent’s ability to recognize and represent object categories,
abstract concepts, and semantic properties. It enables the agent to understand what an object is, what category it belongs to, and what attributes it possesses. 
The interrelations among these three types of cognition are illustrated in Fig.\ref{three-basic-cognition}.

Note that we incorporate these three dimensions because semantically intelligent behavior arises from their integration rather than from any single dimension in isolation. While current embodied agents often address temporal reasoning for task planning, spatial understanding for navigation, and conceptual knowledge for object recognition as separate modules, this disjoint treatment leads to fragmented and shallow semantic understanding~\cite{thomason2020vision,langley2025spatial}.
For example, when an embodied agent is instructed to \textit{``move the fragile vase carefully to the left shelf,''} it must integrate temporal cognition (understanding the need for careful, controlled movement over time), spatial cognition (representing the current location, path to the left shelf), and conceptual cognition (understanding what ``fragile'' implies about handling requirements).
The framework is extensible and allows for the integration of additional cognitive dimensions, depending on the specific application needs or target domains.

As depicted in Fig.~\ref{fig-framework}, the architecture comprises three principal processing layers organized in increasing levels of abstraction: (1) semantic perception, (2) semantic reasoning, and (3) semantic cognition. 
These layers operate under the regulatory influence of a (4) metacognition process that spans all levels. 

\noindent \textbf{(1) Semantic Perception:} The semantic perception layer processes input from multiple sensory modalities (shown at the bottom of Fig.~\ref{fig-framework}) through three parallel pathways corresponding to the fundamental dimensions of cognition.
These pathways implement a computational analog of the FIT psychological model ~\cite{aggelopoulos2015perceptual}, where separate feature dimensions are processed in parallel before being integrated through attentional mechanisms.
The embodied agent typically employs a diverse sensor array to capture various aspects of the environment, including RGB-D cameras for visual and depth information, microphones for audio input, inertial measurement units (IMUs) for proprioception, tactile sensors for surface properties and force feedback, and infrared sensors for thermal detection. These multimodal inputs are processed by modality-specific encoders and then assigned to the appropriate perception pathways: temporal, spatial, or conceptual. This targeted assignment ensures that each cognitive pathway receives structured and relevant inputs, allowing the agent to construct semantically meaningful representations grounded in rich, multimodal observations.

\begin{itemize}
    \item \textbf{Temporal perception} processes basic time-based features, including Events (discrete occurrences), Durations (temporal extents), and Intervals (time between events). This pathway implements computational mechanisms inspired by research on human temporal cognition~\cite{grondin2010timing}, enabling the agent to detect temporal patterns critical for understanding dynamic environments.
    \item \textbf{Spatial perception} extracts basic spatial features, including Distance (metric separation between entities), Direction (orientation relationships), and Location (position within reference frames) relationships between objects. This pathway draws on spatial cognition theories~\cite{montello1998new}, particularly those concerning egocentric (self-centered) and allocentric (environment-centered) spatial representation.
    \item \textbf{Conceptual perception} identifies intrinsic object properties, including Material (substance composition), Shape (geometric form), and Function (usage affordances). This pathway implements computational analogs of perceptual categorization mechanisms~\cite{nosofsky2002exemplar,tiganj2025comparing}, enabling the recognition of semantic properties directly from sensory data.
\end{itemize}

The integration of these perceptual dimensions occurs through a cross-attention mechanism inspired by FIT~\cite{aggelopoulos2015perceptual}, binding basic features across dimensions into coherent object and event representations. This mechanism is formalized as:
\begin{equation}
    \mathbf{O} =Bind(Attend(\mathbf{F_t}, \mathbf{F_s}, \mathbf{F_c})).
\end{equation}
where $\mathbf{F_t}, \mathbf{F_s}$, and $\mathbf{F_c}$ represent basic temporal, spatial, and conceptual feature sets, respectively.
For example, consider an agent observing a person pouring water from a bottle into a glass. The temporal perception pathway extracts event features (the discrete action of tilting the bottle), duration features (how long the pouring continues), and interval features (the timing between the start of pouring and when the water reaches the glass). Through attention and binding mechanisms, these temporal features are integrated to represent the sequential order of the pouring event—beginning, continuous flow, and termination.
Similarly, the spatial perception pathway extracts distance features (the separation between bottle and glass), direction features (the orientation of the bottle relative to the glass), and location features (positions of both objects in the environment). When bound together through cross-attention, these spatial features enable the agent to perceive not just static spatial relationships but also the motion trajectory of the water flowing from bottle to glass.

From the perspective of information flow, the output of the semantic perception layer serves two distinct yet complementary functions within the cognitive architecture. One part of the integrated perceptual representation is routed directly into the agent’s working memory as structured input, enabling immediate use in downstream modules such as decision-making, planning, or reactive control. The other part is propagated upward to the semantic reasoning layers, where it undergoes further abstraction and relational inference to support the extraction of higher-order semantic structures. This dual-pathway design allows the agent to maintain both low-level perceptual grounding and high-level semantic understanding, facilitating context-sensitive and semantically coherent behavior.

\noindent \textbf{(2) Semantic Reasoning:} The semantic reasoning layer implements structured inference mechanisms that transform integrated percepts into semantic knowledge. 

\begin{itemize}
    \item In \textbf{temporal reasoning}, it incorporates axioms about temporal relations, such as the transitivity of time flow:
\begin{equation}
   \forall x \forall y \forall z(x \prec y \wedge y \prec z \rightarrow x \prec z). 
\end{equation}
This allows the agent to infer temporal relations between events that were not directly observed together. Based on perceptual data and reasoning rules, such as instant-based models that represent time as discrete moments and interval-based models that treat time as continuous periods, the agent can infer additional semantic knowledge, thereby significantly enriching its understanding of the physical world.
As shown in Fig.~\ref{fig-framework}, the temporal reasoning can perform dependency reasoning and temporal prediction. Dependency reasoning identifies structural or causal relationships between events or states. For instance, if the agent perceives that a cup has been knocked over, it can infer that the liquid inside is likely to spill, formalized as:
\begin{equation}
    Cup(x) \wedge KnockOver(x)  \rightarrow Spill(y).
\end{equation}
This enables the agent to anticipate consequences and plan accordingly. Temporal prediction, on the other hand, leverages observed sequences to estimate future states:
\begin{equation}
    P(e_{t+1} \mid e_t, e_{t-1}, \dots, e_1).
\end{equation}
where $e_t$  denotes an event at time $t$. This predictive capacity allows the agent to proactively adjust its behavior based on likely developments in its environment, enhancing safety, efficiency, and task fluency.
\item For \textbf{spatial reasoning}, two representative forms of inference are relation inference and collision detection.
Specifically, relation extraction involves reasoning about spatial relationships between entities in the environment, such as ``left of,'' ``inside,'' ``on top of,'' or ``near.''  For example, suppose we know that the vase is on the table, the table is to the left of the bed, and the bed is near the window. From these relationships, we can infer that the vase is to the left of the bed, but we cannot directly infer that the vase is near the window unless we explicitly define the relationship between the table and the window. The relations can be encoded using spatial logic or graph structures, for example:
\begin{align}
    \text{OnTopOf}(Vase, Table) \wedge \text{LeftOf}(Table, Bed) \rightarrow \text{LeftOf}(Vase, Bed).
\end{align}
Inferring such relationships allows the embodied agent to dynamically update and refine its spatial understanding of the environment, supporting tasks like navigation, object localization, and interactive reasoning.
Collision detection refers to the agent's ability to predict and avoid spatial conflicts with objects in its environment. By modeling object trajectories and maintaining up-to-date spatial maps, the agent can formally reason about potential intersections between its own path and surrounding entities. This can be done by checking if the distance between the agent and an object is smaller than a certain threshold, indicating a potential collision:
\begin{align}
    \text{Distance}(P_{\text{agent}}(t), P_{\text{object}}(t)) < \epsilon \rightarrow \text{CollisionRisk}(t).
\end{align}
where \(P_{\text{agent}}(t)\) and \(P_{\text{object}}(t)\) represent the predicted positions of the agent and object at time \(t\), and \(\epsilon\) is a threshold value that defines the collision zone. This capability is critical for safe and efficient navigation in dynamic environments.
\item For \textbf{conceptual reasoning}, the agent can perform both role reasoning and functional reasoning. Role reasoning focuses on inferring the social or contextual roles that an entity plays based on perceptual and situational cues. For instance, an agent may distinguish whether a person is acting as a teacher or a nurse by integrating observations such as attire, location, and associated activities.
Functional reasoning, on the other hand, enables the agent to infer an object's intended or most likely use based on its perceived properties and context. For example, determining whether an object is food or decoration might involve reasoning over features such as shape, color, material, and location:
\begin{align}
    \text{IsEdible}(y) \wedge \text{LocatedIn}(y, \text{kitchen}) \rightarrow \text{Function}(y) = \text{food}.
\end{align}
These reasoning capabilities are essential for embodied agents to interpret abstract semantic roles and functions, thereby enabling goal-directed behavior, human-like understanding, and task-relevant generalization across contexts.
\end{itemize}

\noindent \textbf{(3) Semantic cognition:} The semantic cognition layer integrates the outcomes of temporal reasoning, spatial reasoning, and conceptual reasoning into a unified cognitive representation. 
The integration of these three cognitive dimensions occurs through a multi-level aggregation process that establishes cross-dimensional semantic correspondence~\cite{ralph2017neural}.
Formally, this integration can be modeled as a semantic tensor:
\begin{equation} \mathcal{K} = \mathcal{T} \otimes \mathcal{S} \otimes \mathcal{C}.
\end{equation}
where $\mathcal{K}$ represents the unified semantic tensor resulting from the aggregation of the temporal ($\mathcal{T}$), spatial ($\mathcal{S}$), and conceptual ($\mathcal{C}$) tensors, each encoding information from their respective cognitive dimensions.

A feasible approach for aggregating semantic cognition is through graph-based integration, where each dimension, such as temporal, spatial, and conceptual, is represented as a node in a graph. Edges between these nodes capture the relationships or dependencies between different dimensions. This allows the agent to combine information from various sources and reason about complex, higher-level semantic structures.
Another method is probabilistic reasoning, where techniques like Bayesian networks or belief propagation allow the agent to aggregate uncertain or incomplete information from multiple dimensions, improving the coherence and robustness of its semantic knowledge. This approach helps the agent synthesize knowledge from different sources and make more informed decisions.

Overall, this integrated semantic cognition enables complex understanding that transcends any single dimension, such as recognizing that \textit{``a cup of hot coffee placed near the edge of a table where children are playing''} represents a potential hazard—an inference that requires temporal understanding (children's movements over time), spatial awareness (proximity to edge), and conceptual knowledge (properties of hot liquids and fragile containers). The resulting unified semantic representation is dynamically updated based on new perceptual input and reasoning outputs, and provides the foundation for the agent's decision-making processes by populating working memory with contextually relevant semantic knowledge.

\noindent \textbf{(4) Metacognition Process:} 
As illustrated in Fig.~\ref{fig-framework}, the metacognition process component spans vertically across all three layers of the semantic cognitive architecture, implementing regulatory mechanisms that monitor, evaluate, and control semantic processing. This metacognitive system, inspired by theories of human metacognition  mechanisms~\cite{wang2023metacognitive,fleming2024metacognition}, provides critical self-regulatory capabilities that enhance the robustness and adaptability of semantic intelligence.
It is composed of two key sub-modules: self-monitoring and self-regulation.

\begin{itemize}
    \item \textbf{Self-monitoring} tracks the internal information flow across 
semantic perception, reasoning, and cognition, identifying internal 
inconsistencies, task execution failures, and unexpected environmental 
feedback. For example, it may detect temporal mismatches between predicted 
and observed events, or conflicts between expected and actual spatial 
configurations. Operating continuously across all layers, this module 
maintains the agent's awareness of its cognitive state and alerts the 
system to potential errors in real-time.
    \item \textbf{Self-regulation} acts upon detected inconsistencies by adjusting 
cognitive strategies and initiating compensatory actions. It can reallocate 
attentional focus across temporal, spatial, and conceptual dimensions, 
select alternative reasoning pathways, or retrieve more contextually 
relevant semantic knowledge from memory. This process mirrors the principle 
of top-down modulation in biological cognition~\cite{gilbert2013top}, where 
higher-level control systems influence perceptual and integrative functions 
based on goals and contextual constraints. Through continual regulation, 
the agent maintains semantic coherence and adaptability in complex, 
dynamic environments.
\end{itemize}

% Basic Ability definition

\subsection{Semantic-Driven Decision}

% Decision Process

\begin{figure*}[h]
    \includegraphics[width=\linewidth]{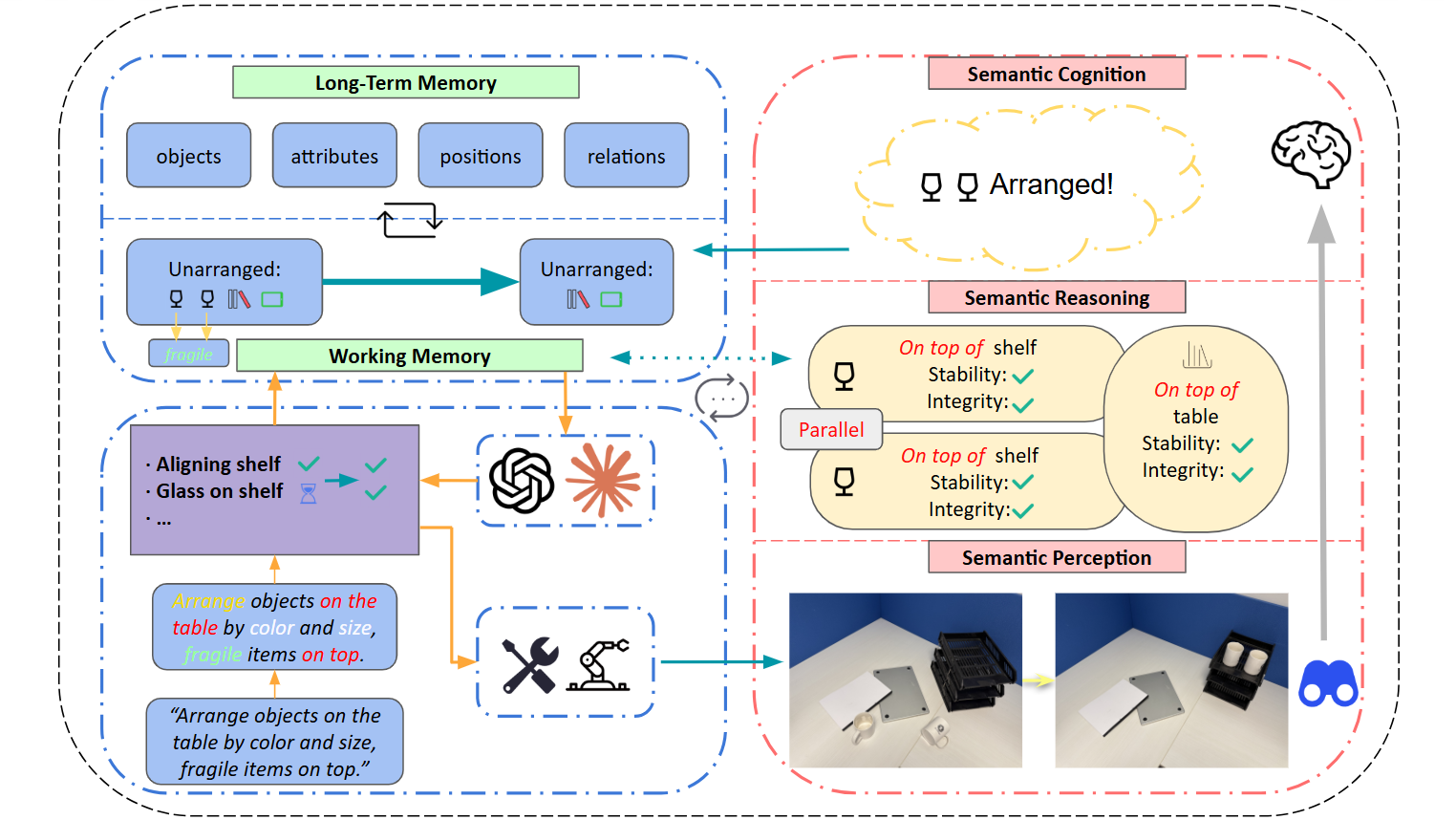}
    \caption{An example of semantic-driven decision process.}
    \label{fig-example}
    \vspace{-15pt}
\end{figure*}

We hereby propose the semantic\mbox{-}driven decision loop within the SIDE agent framework as illustrated in the left panel of Fig.~\ref{fig-framework}. This structure enables agents to process complex tasks by grounding every action in deep contextual semantics and prior knowledge. At the core of this process is the tight coupling between semantic memory\mbox{-}a structured repository of generalizable knowledge and concepts acquired over time\mbox{-}and the mechanisms of working memory and planning, producing a system capable of nuanced, adaptive, and context-sensitive behavior. To better elucidate the loop process, Fig.~\ref{fig-example} provides a specific example delineating the cyclical stages and their interdependencies between the semantic cognitive architecture.

% task specification

The process commences with the interpretation of a task specification. 
% The agent should construct a conprehensive instruction including intent, constraints and desired outcomes by linking the linguistic of symbolic input with its internal semantic schema, which actually demands highly on its semantic understanding ability of language and prior environmental knowledge. 
The agent should construct a comprehensive internal instruction that encompasses the intent, relevant constraints and desired outcome of the task. This instruction requires mapping the linguistic or symbolic input onto the agent's internal semantic schema, which highly demands on the agent's capability to understand language and apply prior environmental knowledge. 
% example
For example, consider the task specification shown in Fig.~\ref{fig-example}:\textit{“Arrange the objects on the table according to both color and size, ensuring fragile items are placed on top.”} Here, the agent must not only parse the explicit requirements\mbox{-}identifying ``objects,'' understanding the dual criteria of ``color and size,'' and recognizing the special handling constraint for ``fragile items''\mbox{-}but also resolve potential ambiguities, such as the order of sorting and the definition of ``top'' in the current context. In forming the task instruction, the agent leverages semantic knowledge of how objects are typically categorized by color and size, as well as task\mbox{-}specific knowledge about handling fragile objects.

% planning with working memory(semantic)

Upon the semantic representation of the task, the agent proceeds to the planning phase, which constitutes a core cognitive module and is closely interacted with working memory. Working memory is conceptualized as an active buffer that holds the immediate context of reasoning: the newly\mbox{-}formed semantic instruction, recent sensory inputs from the environment and a subset of semantic knowledge most pertinent to the task at hand. The interplay between working memory and long-term memory is dynamic; as planning proceeds, working memory activates the long\mbox{-}term memory, selectively retrieves useful experience of previous decision cycles from episodic memory and refreshes the agent's semantic cognition of the current world. Importantly, as the agent acts and perceives, new experiences are encoded from working memory and integrated into long\mbox{-}term memory, effectively refining the agent’s knowledge base through continual learning.  %新形成cognition,与semantic memory交互 by representation.

% LLM interaction

A crucial component of the semantic-driven decision procedure is how planning is facilitated via LLMs, VLMs, or VLAs. With comprehensive task semantics and up\mbox{-}to\mbox{-}date context in the working memory, the agent formulates structured queries with precise semantic meaning to the LLM backbone. For instance, the query might take the form: \textit{“Given objects A, B, and C with respective attributes, and noting that object B is marked as fragile, what is an optimal sorting and placement strategy compliant with the instruction?”} The language model responds by synthesizing a plan that draws upon both the agent’s current knowledge\mbox{-}reflected in the query\mbox{-}and broader priors embedded in the model. The output may describe a multi\mbox{-}stage procedure: first clustering objects by color, then sorting within color type by size, followed by delicate placement of fragile items at the top of each stack.The role of the LLM(VLM/VLA) at this stage is to synthesize a response imbued with the agent's knowledge of the current environment, common sense and other aspects stored in the working memory. The agent's planning module then parses the response of LLM which may be in natural language or structured format, extracting key actional guidance or procedural instructions, and decomposes it into actionable commands tailored to the agent’s sensorimotor capabilities.

% loopback

% Given the execution demand of the planning module, the embodied agent tools execute the corresponding order, which is not a simple feedforward process instead. Execution itself becomes an trigger of another decision-making loop:
Execution does not merely follow a feedforward path; rather, it triggers a new cycle of the decision\mbox{-}making loop. As the agent enacts each step, it closely monitors outcomes via both exteroceptive feedback (from external sensors) and proprioceptive feedback (from internal state monitors). Continuing with our example, suppose the agent picks and places a blue, fragile cup atop a stack of blue plates. If the stack wobbles, the proprioceptive sensors may indicate instability, prompting a real\mbox{-}time re\mbox{-}evaluation of the plan. This ongoing feedback is compared with the expected results elaborated in the semantic model of the original task. Any mismatch or unexpected contingency, such as an object being heavier or more fragile than anticipated, activates a new round of updating: working memory is refreshed with the discrepancy, long\mbox{-}term memory is potentially revised with new experience, and new queries are generated for semantic interpretation and planning.

Such cyclical, feedback\mbox{-}driven planning ensures that the agent’s cognition remains adaptive and grounded in up\mbox{-}to\mbox{-}date, context\mbox{-}rich semantics. The semantic\mbox{-}driven decision loop is, therefore, deeply layered: it begins with the semantic decoding of task intent, proceeds through a dynamic interplay between working memory and episodic/semantic long\mbox{-}term memory, leverages powerful LLMs or VLMs for inference and planning, and grounds all outputs in actionable instructions whose efficacy is verified through execution and feedback. Should new constraints or failures be encountered, the entire cycle restarts with enriched cognition.

% conclusion

In summary, the proposed semantic\mbox{-}driven decision loop offers a robust and extensible pipeline for embodied agents. By synthesizing the superior generalization capabilities of modern LLMs with the structured, context\mbox{-}aware reasoning afforded by semantic memory architectures, our framework enables agents not only to precisely understand and execute complex instructions, but also to continuously shape and refine their cognitive models of the environment at a semantic level.
% To summarize, the semantic-driven decision process is a deeply layered loop. It starts with the semantic decoding of task intent, proceeds through the dynamic update mechanism between working memory and the agent's long-term memory including episodic and semantic parts, leverages state-of-the-art language models for context-sensitive planning, and grounds the output actionable instructions into execution, where the performance feedback is collected for cognition update, and the next cycle of decision is triggered. The strength of this pipeline lies in its ability of merging the LLMs that have advanced generalization ability with the reliability of structured semantic interpretation, yielding agents capable of precisely understanding and shaping its cognition towards the environment in semantic level.

% perceptual input parsing + semantic-driven memory update mechanism + hierarchical reasoning + self-reflection\&adaption

\section{Potential Applications}
\noindent\textbf{Navigation with Semantic Intelligence}: The SIDE framework enables semantically informed navigation by integrating spatial, temporal, and conceptual understanding throughout the perception, reasoning, and cognition layers. Unlike traditional methods that rely solely on geometric mapping and planning, agents equipped with SIDE can recognize functionally meaningful regions, anticipate dynamic changes, and interpret high-level contextual cues. For example, when navigating in an indoor environment, the agent can infer that a kitchen is typically near a dining area, or that the presence of a sleeping person indicates the need to select a quieter route.
Through semantic perception, the agent identifies relevant environmental elements such as objects, rooms, and people. Semantic reasoning allows it to interpret relationships and make predictions about scene dynamics, while semantic cognition supports goal selection and long-horizon planning under context constraints. When failures occur, such as navigation delays or localization errors, the metacognitive module detects the anomaly and activates self-regulation to reallocate attention, revise the plan, or retrieve context-appropriate knowledge. This enhances the robustness, adaptability, and contextual awareness of the agent’s navigation behavior.

\noindent\textbf{Manipulation with Semantic Intelligence}: The SIDE framework provides a foundation for context-aware and semantically informed manipulation by integrating perceptual input with high-level reasoning and goal-directed cognition. Unlike traditional manipulation approaches that rely primarily on geometric features and low-level control, SIDE enables agents to interpret the functional roles of objects, understand task-specific constraints, and reason about abstract goals. For example, in a domestic tidying task, the agent can recognize that a cup placed on the floor violates contextual expectations and infer that its appropriate location is likely within the kitchen environment.
At the perceptual level, the agent extracts semantic attributes such as object category, state (e.g., clean or dirty), and physical affordances. Through semantic reasoning, it infers relational structures and task-consistent spatial configurations, allowing for more purposeful interactions with the environment. Semantic cognition supports hierarchical planning, enabling the agent to sequence actions in alignment with broader task goals. When manipulation failures or contextual inconsistencies arise, the metacognitive component monitors the execution process, evaluates deviations from expected outcomes, and initiates regulatory responses. These include strategy adjustment, attentional reallocation, or retrieval of alternative semantic knowledge, thereby enhancing robustness and adaptability in complex, real-world manipulation tasks.

\section{Future  Directions}
% \section{Building SIDE Agents}

In this section, we identify key research opportunities to advance semantic intelligence in embodied agents. As the development of semantic intelligence continues to progress, there are several critical areas that require focused research efforts. These opportunities not only build on the theoretical foundation established by the SIDE framework but also aim to enhance the practical capabilities of embodied agents, enabling them to more effectively understand, reason, and adapt in dynamic environments.

\noindent\textbf{Testing}: The SIDE agent framework introduces core capabilities that merit rigorous empirical validation. Future studies should investigate the framework’s ability to integrate multi-dimensional semantic information, specifically temporal, spatial, and conceptual elements, under challenging perceptual conditions such as occlusion, noise, or environmental variability. Each of the three specialized reasoning pathways warrants individual evaluation; temporal prediction, spatial relation extraction, and conceptual categorization tasks can be used to assess their respective effectiveness. In addition, the framework’s integrated cognition mechanisms, particularly its use of aggregation operations to construct unified semantic representations, should be examined for their capacity to support cross-dimensional reasoning and generalization to unfamiliar scenarios. The metacognitive regulation component presents additional research potential. Its self-monitoring and self-regulation functions should be evaluated for their roles in managing uncertainty, reallocating cognitive resources, and supporting goal-directed information seeking under varying task conditions.

\noindent\textbf{Enhancement}: 
To enhance embodied agents based on the SIDE framework, future research should focus on addressing specific capabilities that agents may lack. This involves leveraging the framework’s principles to enhance the agents' ability to integrate multi-dimensional semantic information. For example, multimodal learning techniques can be applied to align temporal, spatial, and conceptual representations, while curriculum learning strategies could support the gradual acquisition of semantic competencies. In cases where agents do not possess sufficient social and cultural reasoning, integrating models of social norms, conventions, and theory of mind mechanisms will be essential for improving their understanding of human intentions and enabling more sophisticated human-agent interactions. Furthermore, improving computational efficiency through neuromorphic computing techniques, which mimic biologically inspired architectures, can enhance event-driven processing in agents. Finally, strengthening cross-modal grounding mechanisms will allow agents to better connect abstract concepts with perceptual experiences, enhancing their semantic understanding.

% \section{Evaluating SIDE Agents}

% \section{Discussion}

\section{Conclusion}
In this paper, we emphasize the critical role of semantic intelligence in enabling embodied agents to operate effectively in complex, real-world environments. We introduce the Semantic Intelligence-Driven Embodied (SIDE) agent framework, which integrates temporal, spatial, and conceptual dimensions of semantic understanding inspired by principles from cognitive science. To implement the framework, we design a semantic cognitive architecture and a semantic-driven decision loop. We highlight two representative applications, navigation and manipulation, to demonstrate the potential of the SIDE framework in enhancing embodied agent performance. Furthermore, we outline several promising future research directions grounded in the framework’s core capabilities. 
% The SIDE agent framework sets the foundation for developing agents that can truly understand their environment and adapt in human-like ways.

% \clearpage
\bibliographystyle{ACM-Reference-Format}
\bibliography{sample-base}


% Machine Memory Intelligence: Inspired by Human Memory Mechanisms

\end{document}